# MultiLoad-GAN: A GAN-Based Synthetic Load Group Generation Method Considering Spatial-Temporal Correlations

Yi Hu, *Student Member, IEEE*, Yiyan Li\*, *Member, IEEE*, Lidong Song, *Student Member*, *IEEE*, Han Pyo Lee, *Student Member,* PJ Rehm, Matthew Makdad, Edmond Miller*, and* Ning Lu, *Fellow, IEEE*

*Abstract*— **This paper presents a deep-learning framework, Multi-load Generative Adversarial Network (MultiLoad-GAN), for generating a group of synthetic load profiles (SLPs) simultaneously. The main contribution of MultiLoad-GAN is the capture of spatial-temporal correlations among a group of loads that are served by the same distribution transformer. This enables the generation of a large amount of correlated SLPs required for microgrid and distribution system studies. The novelty and uniqueness of the MultiLoad-GAN framework are three-fold. First, to the best of our knowledge, this is the first method for generating a group of load profiles bearing realistic spatial-temporal correlations simultaneously. Second, two complementary realisticness metrics for evaluating generated load profiles are developed: computing statistics based on domain knowledge and comparing high-level features via a deep-learning classifier. Third, to tackle data scarcity, a novel iterative data augmentation mechanism is developed to generate training samples for enhancing the training of both the classifier and the MultiLoad-GAN model. Simulation results show that MultiLoad-GAN can generate more realistic load profiles than existing approaches, especially in group level characteristics. With little finetuning, MultiLoad-GAN can be readily extended to generate a group of load or PV profiles for a feeder or a service area.**

## I. INTRODUCTION

SMART meter data are essential in power distribution system analysis, for instance, modeling load behaviors, conducting renewable integration studies, and developing demand response programs. However, due to security and privacy considerations, utilities cannot share a large amount of smart meter data with the research community for carrying out such analysis. Therefore, using *synthetic load profiles* (SLPs) derived from smart meter data becomes an increasingly attractive solution.

SLPs are generated load profiles bearing similar characteristics as the real ones. In general, there are two approaches for generating SLPs: *simulation-based* and *data-driven*. Table I summarizes the advantages and disadvantages of the state-of-the-art SLP generation methods and compares our algorithm with the existing ones. As can be seen in the table, up till now, all existing generative methods generate SLPs one at a time. There is no generative method proposed for generating a group of SLPs served by the same distribution transformer or the same feeder, where the SLPs have strong spatial-temporal correlations.

Note that such spatial-temporal correlations exist because consumers at the same geographical location experience similar weather conditions and share similar demographical characteristics (e.g., house type, income level, and living pattern), making the weather-dependent loads and consumption patterns have similar variations. Therefore, the time-series load profiles served by the same transformer or feeder exhibit distinct group-level characteristics. By simply selecting load profiles randomly from a database or generating load profiles for each user one at a time to form a load group, one cannot capture such group-level spatial-temporal correlations. Some forecasting methods based on copula select load profiles by influence factor to form a group, as shown in previous studies [13][14]. These methods typically rely on historical data and involve a two-step forecasting and selecting process. However, copula alone cannot fully capture the spatial and temporal correlations between loads served by the same distribution transformer or feeder. This is because high-level hidden features are usually difficult to explicitly formulate and can only be learned using Deep Learning techniques.

To bridge this gap, we propose a deep-learning framework, called the Multi-load Generative Adversarial Network (MultiLoad-GAN), to generate a group of SLPs simultaneously. The contributions are three-fold. *First*, MultiLoad-GAN captures the spatial-temporal correlations among loads in a load group to enable the generation of correlated realistic SLPs in large quantity for meeting the emerging need in microgrid and

This material is based upon work supported by the U.S. Department of Energy's Office of Energy Efficiency and Renewable Energy (EERE) under the Solar Energy Technologies Office. Award Number: DE-EE0008770.

Yi Hu, Lidong Song, Han Pyo Lee, and Ning Lu are with the Electrical & Computer Engineering Department, Future Renewable Energy Delivery and Management (FREEDM) Systems Center, North Carolina State University, Raleigh, NC 27606 USA. (yhu28@ncsu.edu, lsong4@ncsu.edu, hlee39@ncsu.edu, nlu2@ncsu.edu).

Yiyan Li (corresponding) is with the College of Smart Energy, Shanghai Non-Carbon Energy Conversion and Utilization Institute, Shanghai Jiao Tong University, Shanghai, 200240, China, and also with the Electrical and Computer Engineering Department, Future Renewable Energy Delivery and Management Systems Center, North Carolina State University, Raleigh, NC 27606 USA (email: yiyan.li@sjtu.edu.cn).

PJ Rehm is with ElectriCities. Matthew Makdad and Edmond Miller are with New River Light and Power.

distribution system planning. This is achieved by the novel profile-to-image encoding-decoding method and the MultiLoad-GAN architecture design. *Second*, two complementary metrics for evaluating realisticness of generated load profiles are developed: computing statistics based on domain knowledge and comparing high-level features via a deep-learning classifier. *Third*, to tackle data scarcity, a novel iterative data augmentation mechanism is developed to generate training samples for enhancing the training of both the deep-learning classifier and the MultiLoad-GAN model, which can improve the performance of MultiLoad-GAN by approximately 4.07%.

TABLE I
COMPARISON OF OUR MULTILOAD-GAN MODEL WITH STATE-OF-THE-ART GENERATIVE METHODS

| | | Description | Advantages | Disadvantages | Model output |
|---|---|---|---|---|---|
| Model-based methods [1][2] | | Use physical models, such as building thermodynamics and customer behavioral models, to simulate electricity consumption profiles. | Explainable as the models reflect the laws of physics when describing the behavior behind field measurements | Require detailed physics-based models with many inputs and require parameter tuning. | **Single load profile** (When generating a load profile, the methods do not consider the spatial-temporal correlations among a group of generated load profiles) |
| Data-driven methods | Clustering based [3][4] | Cluster existing load profiles into different categories so that by combining the load profiles across different categories, SLPs are generated. | Easy to implement and can represent some realistic load profile characteristics. | Lack of diversity when using combinations of a limited number of existing profiles. | |
| | Forecasting based [5]-[8] (the benchmark method) | Generate SLPs based on publicly available load or weather data. | Easy to implement and flexible to generate load profiles with different lengths and granularities. | Depend heavily on historical data. The generated load profiles have similar patterns with historical data, therefore, lack of diversity. | |
| | **SingleLoad-GAN-based** [10]-[12] (the benchmark method) | GAN-based generative methods to generate the SLP for one customer at a time. | Learn from the real data distribution to generate diversified load profiles with high-frequency details. | Hard to train. | |
| | **MultiLoad-GAN (the proposed method)** | GAN-based generative methods to generate a group of spatial-temporal correlated load profiles simultaneously. Such load profiles can be loads served by the same transformer or feeder. | Learn from the distribution of real data to generate diversified load profiles with high-frequency details. Preserve the spatial-temporal correlations between loads. | Hard to train. | **Multiple spatial-temporal correlated load profiles** |

The rest of the paper is organized as follows. Section II introduces the methodology, Section III introduces the simulation results, and Section IV concludes the paper.

## II. METHODOLOGY

In this section, we first present the terminologies used in the paper and the overall modeling framework. Next, we present MultiLoad-GAN, statistical based evaluation metrics, the training of a classifier to assess the realisticness of generated load profile groups, and the implementation of Automatic Data Augmentation to enhance the algorithm performance.

### A. Terminologies and the Modeling Framework

In this paper, we define **a load group** as loads served by the same transformer. A "***positive sample***" is defined as a group of load profiles from customers supplied by the same transformer. The "***original positive samples***" is the labelled data set given by a utility in North Carolina area including 8 transformers with each serving 8 loads from 2017 to 2020. The "***negative samples***" is the data set that consists of groups of load profiles from customers unlikely to be supplied by the same transformer. This is a unique definition because in power distribution systems, even if a load profile is from a user supplied by another transformer, the load profile is likely to be similar to loads supplied under the same transformer. For example, in a neighborhood, serving which 8 out of 10 neighboring houses is sometimes a random choice by design engineers. In most cases, all $C(10,8)$ combinations can be considered as positive samples. Thus, in the training, the "***original positive samples***" is the ground-truth data set labelled by utility engineers while all negative samples are "***generated negative samples***" generated by us.

Fig. 1 shows the overview of the framework. As shown in Fig. 1(a), the MultiLoad-GAN framework includes three modules: MultiLoad-GAN, Deep-learning classifier (DLC), and Negative Sample Generator (NSG). Initially, because there are no labelled negative samples for training DLC, we develop the NSG module for generating negative samples to enhance the training of the DLC.

As shown in Fig. 1(b), due to security and privacy considerations, the amount of labelled data provided by the utility to train MultiLoad-GAN is usually insufficient. Therefore, to further improve the performance of MultiLoad-GAN, we develop an interactive process, Automatic Data Augmentation (ADA), for generating augmented labelled data, which allows the training of DLC and MultiLoad-GAN to iteratively evolve with the augmented data generation process.

As shown in Fig. 1(c), the realisticness of the generated load groups is evaluated by comparing the generated load groups with the "original positive samples" using two kinds of realisticness metrics: statistics metrics based on domain knowledge and a deep-learning classifier for comparing high-level features.

To the best of our knowledge, there is no other existing approach for generating a group of highly correlated load profiles in the literature. Thus, the goal of our comparison is to demonstrate that when an algorithm generates load profiles one at a time, it cannot generate a group of load profiles that bear correct group-level characteristics. We select SingleLoad-GAN as the benchmark model for performance comparison, because SingleLoad-GAN and MultiLoad-GAN formulate an ablation study in nature. In addition, GAN based models produce more realistic and diversified shape-wise load profile details than the other existing methods by learning the distribution of real data, as shown in [10]. Therefore, the SingleLoad-GAN reproduces the method presented in [10]. Due to differences in input data, SingleLoad-GAN uses different parameters.

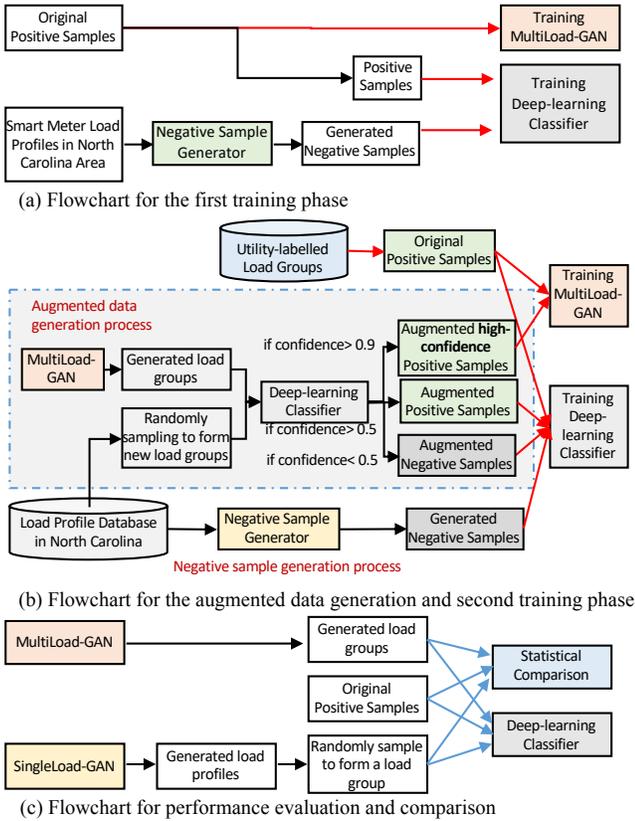

(a) Flowchart for the first training phase

(b) Flowchart for the augmented data generation and second training phase

(c) Flowchart for performance evaluation and comparison

Fig. 1. An overview of the overall modeling framework.

### B. GAN-based Approach

A GAN model consists of two components: a generator network ($G$) and a discriminator network ($D$). A latent vector $\mathbf{z}$, usually a Gaussian noise, is used as the input to generate the target output $G(\mathbf{z})$. Then, the generated data $G(\mathbf{z})$ and the real data $\mathbf{x}$ are sent to $D$. The goal of $D$ is to distinguish which data samples are real and which are fake.

The training of a GAN model is an alternative and adversarial process: $G$ tries to generate samples $G(\mathbf{z})$ that can fool $D$; $D$ learns to distinguish between $G(\mathbf{z})$ and $\mathbf{x}$ by assigning larger probabilities to $\mathbf{x}$ and smaller ones to $G(\mathbf{z})$. As introduced in [9], this process is formulated as a minimax game

$$\min_G \max_D \left[ E_{\mathbf{x} \sim P_r}[\log(D(\mathbf{x}))] + E_{\hat{\mathbf{x}} \sim P_g}[\log(1 - D(\hat{\mathbf{x}}))] \right] \quad (1)$$

where $P_r$ and $P_g$ are the probability distributions of the training data and the generated data, $E$ is the expectation operator, and $\hat{\mathbf{x}} = G(\mathbf{z})$. According to Wasserstein Generative Adversarial Networks (WGAN) [15][16], the training process will be more stable than the original GAN with the following loss function

$$\min_G \max_{D \in \omega} \left[ E_{\mathbf{x} \sim P_r}[D(\mathbf{x})] - E_{\hat{\mathbf{x}} \sim P_g}[D(\hat{\mathbf{x}})] \right] \quad (2)$$

where $\omega$ is the set of 1-Lipschitz function. A gradient penalty method [16] is proposed to further improve the performance of WGAN. Thus, we adopt the following loss function in our framework

$$
\begin{aligned}
L = {} & E_{\hat{\mathbf{x}} \sim P_g}[D(\hat{\mathbf{x}})] - E_{\mathbf{x} \sim P_r}[D(\mathbf{x})] \\
& + \lambda E_{\hat{\mathbf{x}} \sim P_{\hat{\mathbf{x}}}}[(\|\nabla_{\hat{\mathbf{x}}} D(\hat{\mathbf{x}})\|_2 - 1)^2]
\end{aligned} \quad (3)
$$

where $P_{\hat{\mathbf{x}}}$ is the distribution sampled uniformly along straight lines between pairs of points sampled from the data distribution $P_r$ and the generator distribution $P_g$.

In this paper, we will use SingleLoad-GAN as the baseline model for benchmarking the performance of MultiLoad-GAN, for the reasons explained in section II A. The SingleLoad-GAN based approach is introduced in [10]-[12] and in this paper we reproduced the model in [10].

Let $\mathbf{P}_M^i = [p_1^i, p_2^i, \dots, p_M^i]^T$ represent the load profile of the $i^{\text{th}}$ individual user with $M$ data samples. SingleLoad-GAN generates synthetic profiles, $\hat{\mathbf{P}}_M^i = [\hat{p}_1^i, \hat{p}_2^i, \dots, \hat{p}_M^i]^T$ with similar distribution as $\mathbf{P}_M^i$. After all profiles are generated for $N$ loads, we obtain a load group, $\hat{\mathbf{P}}_{M \times N}$,

$$\hat{\mathbf{P}}_{M \times N} = [\hat{\mathbf{P}}_M^1, \hat{\mathbf{P}}_M^2 \dots, \hat{\mathbf{P}}_M^N] = \begin{bmatrix} \hat{p}_1^1 & \hat{p}_1^2 & \cdots & \hat{p}_1^N \\ \hat{p}_2^1 & \hat{p}_2^2 & \cdots & \hat{p}_2^N \\ \vdots & \vdots & \ddots & \vdots \\ \hat{p}_M^1 & \hat{p}_M^2 & \cdots & \hat{p}_M^N \end{bmatrix} \quad (4)$$

### C. MultiLoad-GAN Model

The configuration of MultiLoad-GAN is shown in Fig. 2. The MultiLoad-GAN generator network is a deep Convolutional Neural Network (CNN). First, a fully connected layer is used to extract features from the input data to a 2D data matrix. Then, transpose convolutional layers with decreasing number of kernels are used to generate load profile groups. ReLU is used as the activation function. Inspired by [17], we use batch normalization following each transpose convolutional layer to enhance the training process. A $Tanh$ layer is added to the end of the generator to normalize the output values into [-1, 1].

The MultiLoad-GAN discriminator is built with a set of convolutional layers with increasing number of kernels. The activation function is Leaky ReLU. Also, a batch normalization layer is added following each convolution layer.

Compared with SingleLoad-GAN, a distinct advancement of MultiLoad-GAN is that it generates $N$ load profiles simultaneously, so we have

$$\ddot{\mathbf{P}}_{M \times N} = \left[\ddot{\mathbf{p}}_M^1, \ddot{\mathbf{p}}_M^2 \dots, \ddot{\mathbf{p}}_M^N\right] = \begin{bmatrix} \ddot{p}_1^1 & \ddot{p}_1^2 & \dots & \ddot{p}_1^N \\ \ddot{p}_2^1 & \ddot{p}_2^2 & \dots & \ddot{p}_2^N \\ \vdots & & \ddots & \vdots \\ \ddot{p}_M^1 & \ddot{p}_M^2 & \dots & \ddot{p}_M^N \end{bmatrix} \quad (5)$$

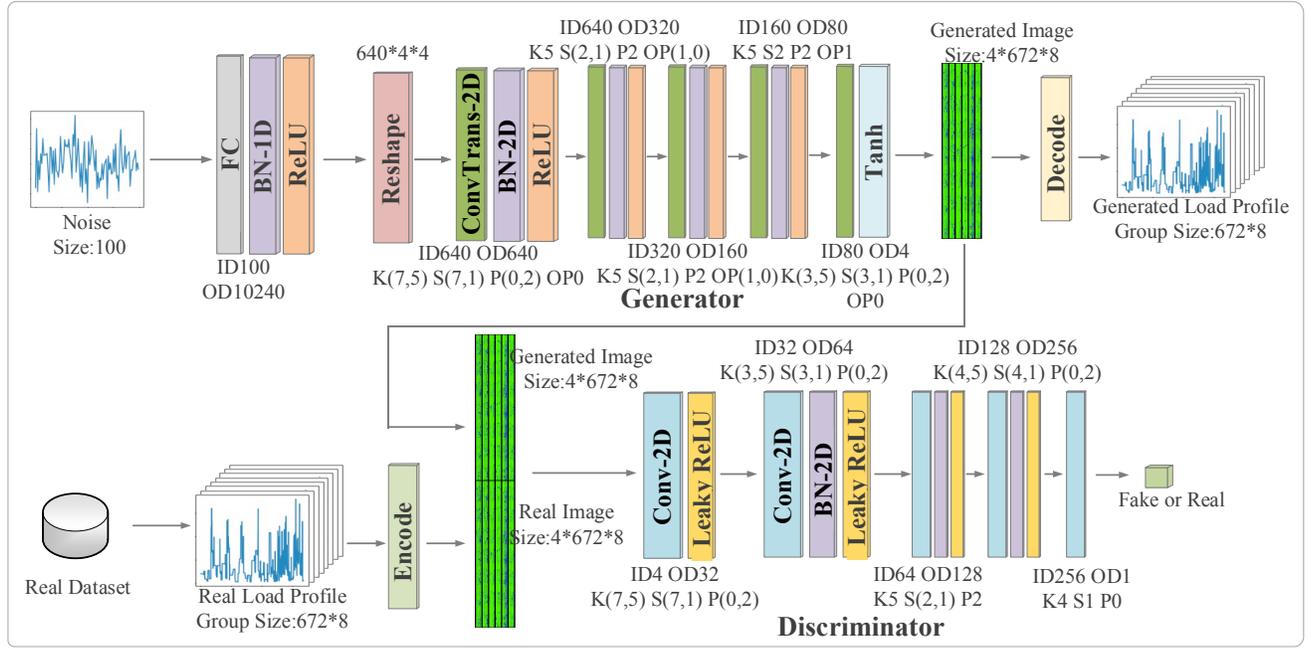

Fig. 2. MultiLoad-GAN architecture with corresponding input dimension (ID), output dimension (OD), kernel size (K), stride (S), padding (P), output padding (OP) for each convolutional layer. The parameter is an example for generating weekly 15-min load group with 8 households.

As shown in Fig. 3, inspired by the image processing encoding process, a unique profile-to-image encoding method is developed to encode a group of load profiles (an $M \times N$ matrix) into 3 color channels (red-R, green-G and blue-B). Dependency of load on temperature is investigated in many load forecasting researchers such as [18][19]. Then, we add a fourth channel to represent temperature ($T$) to reflect the weather dependence.

To encode $\mathbf{P}_{M \times N}$, map $p_m^n \in \mathbf{P}_{M \times N}$ to the RBG channels, $[r_m^n, g_m^n, b_m^n]$, by

$$r_m^n = \begin{cases} 0, & 0 \leq p_m^n < l_1 \\ \dfrac{p_m^n - l_1}{l_2 - l_1}, & l_1 \leq p_m^n < l_2 \\ 1 - \dfrac{p_m^n - l_2}{l_3 - l_2}, & l_2 \leq p_m^n < l_3 \\ 0, & l_3 \leq p_m^n \end{cases}$$

$$g_m^n = \begin{cases} 1 - \dfrac{p_m^n}{l_1}, & 0 \leq p_m^n < l_1 \\ 0, & l_1 \leq p_m^n \end{cases} \quad (6)$$

$$b_m^n = \begin{cases} \dfrac{p_m^n}{l_1}, & 0 \leq p_m^n < l_1 \\ 1 - \dfrac{p_m^n - l_1}{l_2 - l_1}, & l_1 \leq p_m^n < l_2 \\ 0, & l_2 \leq p_m^n \end{cases}$$

$$l_1 = \frac{1}{3} l_3, l_2 = \frac{2}{3} l_3,$$

$$l_3 = \max(p_m^n, \ for \ m \in [0, M], n \in [0, N]).$$

The fourth channel is the temperature channel. Temperature measurement $T_m$ at time $m$ is first normalized by $120°$F in order

to encode it to the brown channel, $[t_m]$, so we have

$$t_m = \frac{T_m}{120} \quad (7)$$

Thus, the load $p_m^n$ and temperature $T_m$ measurement at time point $m$ is encoded into a normalized vector $[r, g, b, t]$ within $[0, 1]$. Then we further convert them into $[-1, 1]$ to benefit the model training process by

$$[r, g, b, t] = \frac{[r,g,b,t] - 0.5}{0.5} \quad (8)$$

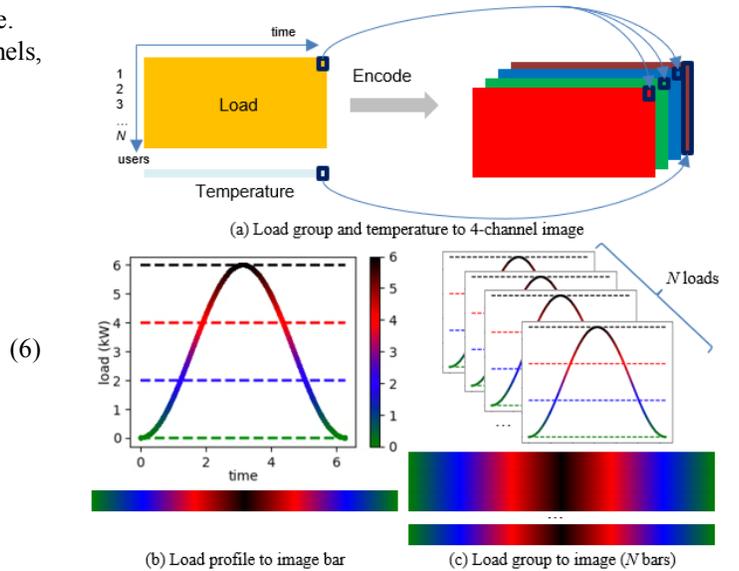

(a) Load group and temperature to 4-channel image

(b) Load profile to image bar

(c) Load group to image ($N$ bars)

Fig. 3. An illustration of the profile–to-image encoding process.

By encoding load profiles into an RBG image, machine learning tools developed in 2D image processing domain can be readily used to extract the spatial-temporal correlations among multiple loads. Also, RGB images make it easier for human to visually recognize load variations patterns. Thus, we consider the profile-to-image encoding method and the corresponding adjustments on the conventional GAN architecture as one of the contributions of the paper.

## D. Realisticness Evaluation

Denote the load group generated by SingleLoad-GAN as $\widehat{\mathbf{P}}_{M \times N} \in \mathbf{\Omega}_{LG}^{SLGAN}$; denote the load group generated by MultiLoad-GAN as $\ddot{\mathbf{P}}_{M \times N} \in \mathbf{\Omega}_{LG}^{MLGAN}$; denote the ground-truth load group as $\mathbf{P}_{M \times N} \in \mathbf{\Omega}_{LG}$. As shown in Fig. 1(c), first, we compare statistical evaluation indices to quantify the realisticness at the household level (i.e., for each individual user) and the aggregation level (i.e., at the transformer level). Ideally, $\ddot{\mathbf{P}}_{M \times N}$ is expected to have similar transformer-level statistics with $\mathbf{P}_{M \times N}$ than those of $\widehat{\mathbf{P}}_{M \times N}$ because spatial-temporal correlations among a group of users are implicitly learned by MultiLoad-GAN. Next, to compare features unable to be captured by human-defined indices, a specialized DLC is trained to assess the realisticness of $\ddot{\mathbf{P}}_{M \times N}$ and $\widehat{\mathbf{P}}_{M \times N}$ by comparing high-level features captured in $\mathbf{P}_{M \times N}$.

### 1) Method 1: Statistical Evaluation

The statistical evaluation metrics are summarized in Table II.

TABLE II
STATISTICAL EVALUATION INDEXES FOR THE $n^{\text{TH}}$ LOAD GROUP

| No. | Indexes |
|-----|---------|
| 1 | Peak load distribution |
| 2 | Mean power consumption distribution |
| 3 | Load ramps distribution |
| 4 | Hourly energy consumption distribution |
| 5 | Daily energy consumption distribution |

First, distributions of each load character index for the generated and ground-truth load groups, $\ddot{\mathbf{P}}_{M \times N}$, $\widehat{\mathbf{P}}_{M \times N}$, and $\mathbf{P}_{M \times N}$, are first calculated at both the household- and transformer- levels. Then, similarities between the distributions of each index for $\mathbf{\Omega}_{LG}^{MLGAN}$, $\mathbf{\Omega}_{LG}^{SLGAN}$ and $\mathbf{\Omega}_{LG}$ are compared to quantify the realisticness of the generated load groups.

### 2) Method 2: Deep-Learning Classification

DLC is trained in parallel with MultiLoad-GAN. DLC can identify real and fake load groups with higher accuracy than the MultiLoad-GAN *discriminator* because both positive and negative samples are used to train DLC, making it a highly "specialized" classifier for identify real and fake load groups.

As shown in Fig. 4, the configuration of DLC includes a deep convolutional network consisting of 5 2-D convolutional layers with increasing number of kernels and 5 fully connected layers with decreasing number of features. The DLC input is a $M \times N$ load group and the output is the probability of realisticness, which reflects how well realistic group-wise spatial-temporal correlations can be captured.

Assume there are Q samples (each sample $\mathbf{P}_{M \times N} \in \mathbf{\Omega}_{LG}$ is a group of load profile with size $M \times N$) used to train the classifier. For the $i^{th}$ sample, the classifier output is

$$C(\mathbf{P}_{M \times N}) = P_{true}(i) \qquad (9)$$

where $P_{true}(i) \in [0,1]$ is the probability for the $i^{th}$ load group to be "real". Thus, we consider the sample to be "positive" if $P_{true}(i) > 0.5$ and "negative" otherwise.

Let $Q_{real}$ be the number of samples classified as "positive".

The Percentage of Real ($POR$) of the dataset is calculated as

$$POR = \frac{Q_{real}}{Q} \times 100\% \qquad (10)$$

Although $POR$ can be used to evaluate the accuracy of the classifier, it cannot reflect the confidence level of the classification results. For example, considering a sample "positive" when $P_{true}(i) = 0.51$ is a less certain judgement than when $P_{true}(i) = 1$. So, we further calculate the Mean Confidence Level of the dataset ($MCL$) as

$$MCL = \frac{1}{Q} \sum_{i=1}^{Q} P_{true}(i) \qquad (11)$$

The similarity of real dataset $\mathbf{\Omega}_{LG}$ and MultiLoad-GAN generated dataset $\mathbf{\Omega}_{LG}^{MLGAN}$ can be calculated by the Fréchet inception distance [20][21] between the two distributions

$$Similarity = FID\left(C(\mathbf{\Omega}_{LG}), C(\mathbf{\Omega}_{LG}^{MLGAN})\right) \qquad (12)$$

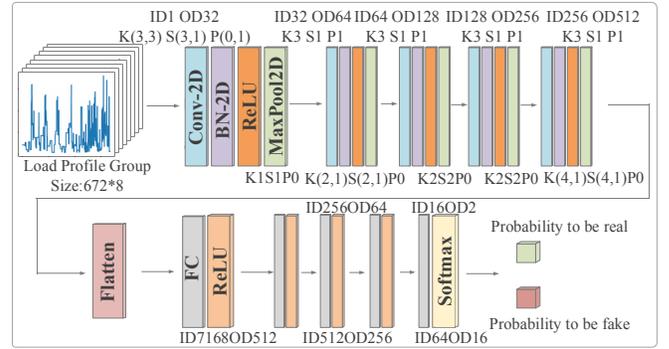

Fig. 4. Classifier architecture with corresponding input dimension (ID), output dimension (OD), kernel size (K), stride (S), and padding (P) for each convolution layer, max pool layer, and fully connected layer. The parameter is an example for generating weekly 15-min load group with 8 households.

## E. Negative Sample Selection

To train DLC, both positive samples (i.e., realistic load groups labeled by 1) and negative samples (i.e., unrealistic load groups labeled by 0) are required. The "original positive samples", $\mathbf{P}_{M \times N}$, is the labelled ground-truth data set. However, because negative samples (i.e., load groups served by different transformers) do not "naturally" exist, they are generated by pulling load profiles from various transformer load groups.

These negative samples play a vital role in preventing bias in load group classification. Without them, the classifier may mistakenly classify all load groups as positive samples, resulting in subpar performance. Therefore, generating high-quality negative samples is crucial for achieving a balanced training dataset, reducing biases, and enabling the classifier to learn how to classify unseen data effectively. This process also enhances the classifier's robustness in the presence of noise or ambiguous cases.

However, negative sample generation is a nontrivial task. Randomly selecting a group of users from a smart meter database that contains load profiles collected in the same area in the same season is a straightforward way to generate a negative sample. However, it is an uncontrolled approach with several drawbacks. First, a significant amount of the generated negative samples are too easy to be classified as "negative".

Thus, DLC cannot learn complex hidden features. Second, an unknown number of the generated negative samples are actually positive samples. This is because, often times, a load served by one service transformer is equally likely to be served by an adjacent transformer with the same size. This phenomenon is quite common when supplying small residential loads. Thus, randomly drawing loads from a regional smart meter database to obtain negative samples is not a reliable negative sample generation strategy.

Therefore, a statistic-based negative sample generation method is developed. First, we obtain operational statistics from the "original positive samples". As shown in Fig. 5(a), we evenly divide the mean power value distribution of the real load profiles into 6 parts. A negative sample can thus be obtained if we randomly select $K$ load profiles from the red box region and $N - K$ load profiles from the black box regions, where $K$ is a random integer in $[0, N/2]$ and randomized in each selection process. By doing so, the obtained load group has a much higher chance to be a negative sample. Similarly, based on Fig. 5(b), we can select negative samples to let the load group having different weekly peak distributions from that of the "original positive samples". By controlling the distance between the "real" and "fake" distributions, one can generate "very negative", "negative", "slightly negative", and "almost positive" samples. This gives the modeler the flexibility to tune the DLC to capture different level of realisticness.

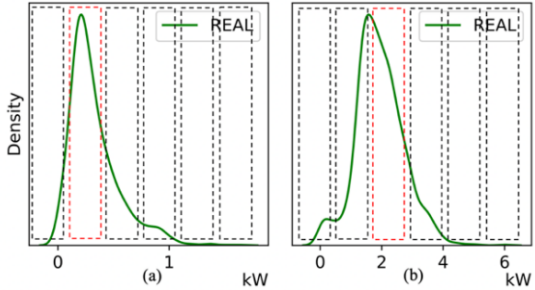

Fig. 5.  (a) Mean power distribution and (b) Peak load distribution.

### F. Automatic Data Augmentation

As shown in Fig. 6, we propose an iterative ADA mechanism that leverages the generation ability of MultiLoad-GAN and the classification capability of DLC to iteratively create augmented training samples in the training (e.g. at step $t - 1$) in order to boost subsequent training (e.g., at step $t$) of both MultiLoad-GAN and DLC. ADA includes three processes: unlabeled data set creation, labelling data for classifier training, and augmented data for MultiLoad-GAN training.

First, we use two methods to create unlabeled load groups, $\Omega_{LG}^{unlabeled}$: 1) using MultiLoad-GAN to generate load groups directly into $\Omega_{LG}^{MLGAN}$, and 2) randomly sampling $N$ load profiles from a smart meter database, $\Omega_{Load}$, to obtain $\Omega_{LG}^{Rand}$. Note that at this stage, $\Omega_{LG}^{unlabeled}$ contains both positive and negative samples.

Next, $\Omega_{LG}^{unlabeled}$ will be labeled by the DLC with parameter, $\theta_C$ obtained from the previous training step, $t - 1$. Note that such labels can include errors, depending on what the accuracy of the DLC is at the training stage. Then, the labeled data, together with the "original positive samples" ( $\Omega_{LG}$ ) and negative samples ($\Omega_{LG}^{Neg}$, selected by NSG in Section II.C) will be used to train the DLC.

Third, once the Classifier is trained, it will immediately be used for identifying positive samples from $\Omega_{LG}^{Rand}$, which will then be used as the augmented dataset $\Omega_{LG}^{Aug}$ for training MultiLoad-GAN. Note that only samples with a high confidence level (e.g., samples with scores > 0.9) will be selected to enhance the quality of the augmented data.

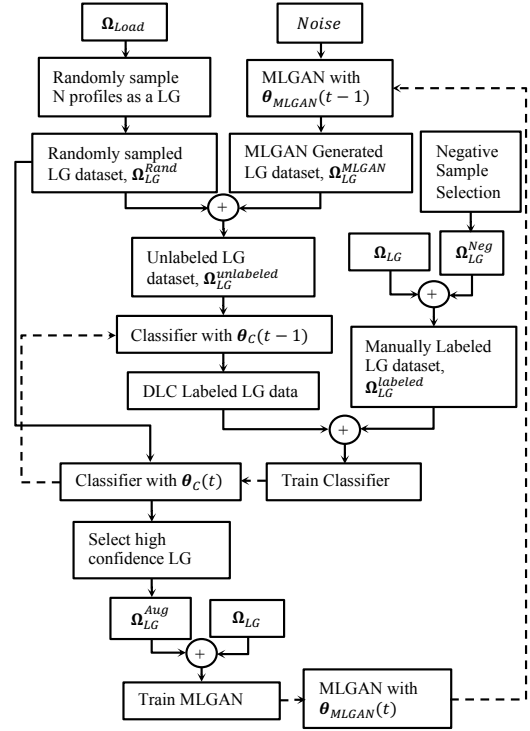

Fig. 6.  Flowchart for the iterative ADA process.

When the interactive training process progresses, the DLC training will improve significantly because it receives an increasing number of high-quality training data generated by MultiLoad-GAN and random sample selection. In return, the classifier can help identify positive samples with higher confidence level to enhance the training of MultiLoad-GAN. The training of MultiLoad-GAN and DLC will be both enhanced until the accuracy of the classifier saturates to a certain level.

## III. SIMULATION RESULTS

In this paper, we use transformer-level load group generation as an example to illustrate the group-load generation process and evaluate algorithm performance.

A utility in the North Carolina area provides us with a labelled data set, which includes 3-year (from 2017 to 2020), 15-minute load profiles collected from 64 residential customers, which are served by eight 50kVA transformers (8 customers per transformer). The corresponding temperature data are from National Oceanic and Atmospheric Administration (NOAA) [22] website. After excluding the missing data, we obtain 1424 load group (each sample is a matrix of $\mathbf{P}_{672 \times 8}$), which are considered as "original positive samples". Thus, the output of the MultiLoad-GAN model should be weekly load profiles (i.e., $M = 4 \times 24 \times 7 = 672$) for a group of 8 loads (i.e., $N$=8) serving by the same transformer.

The power and temperature encoding parameters are given in Table III and the profile-to-image process is illustrated in Fig. 3(b)(c). Hyper-parameter settings of MultiLoad-GAN are given in Table IV. We use the root mean square propagation (RMSProp) optimizer. The model is built in the PyTorch environment and trained on a single NVIDIA GeForce RTX 1080 GPU. Each training takes approximately 2 hours. The architecture of the benchmark model, SingleLoad-GAN is implemented with a set of hyper-parameter settings shown in Table IV. Each training takes approximately 1 hour.

It is important to point out that we do not split the data into training and testing sets for a GAN model, as it is done in other supervised learning. Because the GAN model learns the mapping from the latent vector distribution to the actual data distribution. As a result, the evaluation of the GAN model is not a point-to-point comparison between the generated results and the actual results (because the generated results should be different from any existing ones and therefore is not comparable). Instead, the evaluation focuses on the realisticness of the generated results, such as statistical evaluation, visual inspection, deep learning classification, etc.

### TABLE III
### PARAMETERS USED IN THE PROFILE-TO-IMAGE ENCODING PROCESS

| Load (kW) | Vector [r, g, b] | Temperature (Fahrenheit) | Vector [t] |
|---|---|---|---|
| 0 | [0, 1, 0] | 0 | [0] |
| (0, 2) | g↓, b↑ | | |
| 2 ($l_2$) | [0, 0, 1] | | |
| (2, 4) | b↓, r↑ | (0, 120) | t↑ |
| 4 ($l_3$) | [1, 0, 0] | | |
| (4, 6) | r↓ | | |
| [6 ($l_5$), +∞) | [0, 0, 0] | 120 | [1] |

The loss curves when training MultiLoad-GAN is shown in Fig. 7(a) stage 1. Initially, there is a sharp decrease of the discriminator loss. This means that the discriminator quickly captures the differences between the real ($\mathbf{P}_{672 \times 8}$) and fake ($\tilde{\mathbf{P}}_{672 \times 8}$) load groups generated by the naive generator. When the generator network is stronger than the discriminator network, and able to generate more realistic samples that can fool the discriminator, the loss of the discriminator will increase, and the loss of the generator will decrease. Otherwise, when the discriminator is stronger, the loss of the discriminator will decrease, and the loss of the generator will increase. Such adversarial training process allows both the generator and the discriminator to continuously improve themselves. After about 300 epochs, the generator and discriminator of MultiLoad-GAN reach a balanced state, showing that the generator can generate realistic load groups. The training process of SingleLoad-GAN shown in Fig. 7(b) is similar.

The generated load groups are shown in Fig. 8. It is hard to evaluate the realisticness of a load profile by visually comparing the load profiles, even harder for a human to judge whether a group of load profiles bear similar spatial-temporal correlations. The results show that it is necessary to use statistical metrics and DLC for quantifying realisticness in synthetic load profile generation instead of relying on visual inspection, which is commonly used in image processing domain.

### TABLE IV
### HYPERPARAMETER SETUP FOR THE GAN MODEL

| Parameter | MultiLoad-GAN | SingleLoad-GAN |
|---|---|---|
| Learning rate | 1e-4(D) 1.4e-4(G) | 1e-4(D) 1.2e-4(G) |
| Gradient penalty weight - λ | 10 | 10 |
| Slop of LeakyReLU | 0.2 | 0.2 |
| Batch size | 16 | 64 |
| Training epochs | 300 | 100 |
| Training time | Stage 1: ~2hrs Stage 2: ~4hrs | ~1hr |

#### A. Statistical Evaluation

To compare the performance improvement, we compared 1424 load groups ($\tilde{\mathbf{P}}_{672 \times 8}$) generated by MultiLoad-GAN with 1424 load groups ($\tilde{\mathbf{P}}_{672 \times 8}$) generated by SingleLoad-GAN. By doing so, we have a real load group database ($\mathbf{\Omega}_{LG}$), a MultiLoad-GAN generated load group database ($\mathbf{\Omega}_{LG}^{MLGAN}$), and a SingleLoad-GAN generated database ($\mathbf{\Omega}_{LG}^{SLGAN}$), each having 1424 samples. The load statistics can be calculated at both the household and transformer levels. By comparing the distance between the metric distribution of the generated load groups and the real load groups, we can assess the realisticness of the generated load profiles.

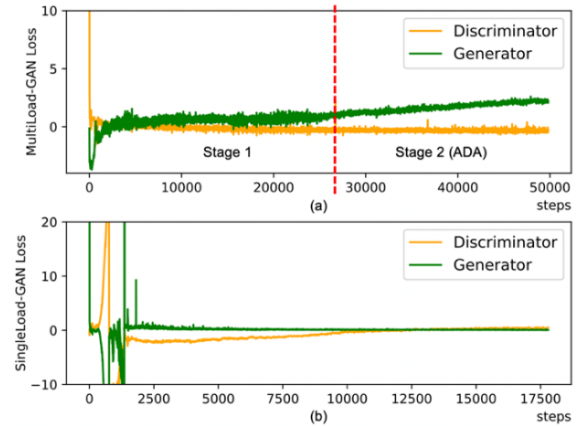

Fig. 7. (a) Loss curves of MultiLoad-GAN, stage 1: Loss curve of discriminator and generator in the MultiLoad-GAN training stage; stage 2: Loss curve of discriminator and generator in the Automatic Data Augmentation (ADA) training stage. (b) Loss curves of SingleLoad-GAN

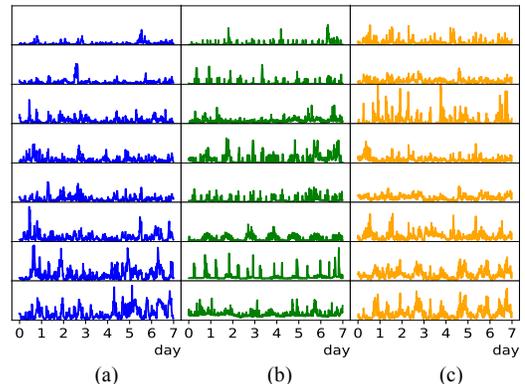

Fig. 8. (a) a MultiLoad-GAN generated load group ($\tilde{\mathbf{P}}_{672 \times 8}$), (b) a real load group ($\mathbf{P}_{672 \times 8}$), (c) a SingleLoad-GAN generated load group ($\tilde{\mathbf{P}}_{672 \times 8}$).

To ensure a comprehensive evaluation, a LSTM-based load forecasting method has been used as an additional benchmark. This method forecasts a set of load profiles using real load groups. The data processing procedure follows a similar approach to SingleLoad-GAN.

### 1) Evaluation at the Household-level

For the household level evaluation, statistics are calculated based on individual load profile. Note that each database contains 1424×8=11392 weekly load profiles.

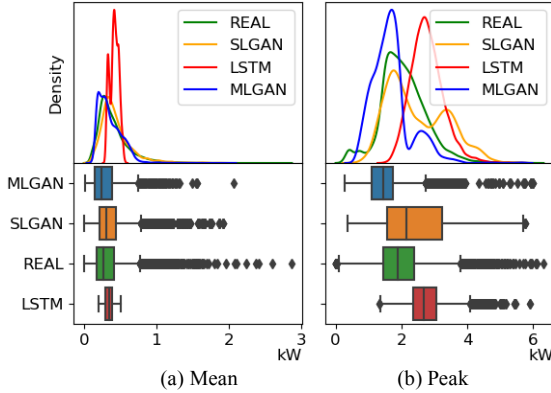

Fig. 9. (a) Household-level mean power distribution curves and boxplots, and (b) Household-level peak load value distributions and boxplots.

- *Mean and peak*. As shown in Fig. 9, MultiLoad-GAN and SingleLoad-GAN can all generate load profiles with the mean value distribution close to that of real load groups. However, SingleLoad-GAN tends to generate load profiles with higher peak values (e.g., from 3 to 5kW), making its peak value distribution deviate from the ground truth. This is because when generating load profiles one at a time, the correlation between users are not considered, making SingleLoad-GAN results less realistic. Due to its nature as a forecasting method relying on historical data, LSTM tends to generate load profiles with narrow ranges for mean and peak values. This limitation makes it challenging to produce diversified load profiles.

- *Load ramps*. The distributions of load ramps on the four data sets are shown in Fig. 10. We can see that MultiLoad-GAN and SingleLoad-GAN show comparable performance on this metric, but LSTM performs worse than them.

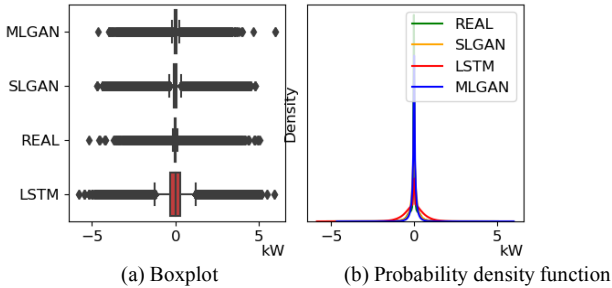

Fig. 10. Household-level load ramp distributions curves and boxplots.

- *Daily and hourly power consumption*. As shown in Figs. 11 and 12, MultiLoad-GAN and SingleLoad-GAN has similar performance on daily power consumption but is slightly worse than SingleLoad-GAN on hourly power consumption. LSTM still perform worse than them.

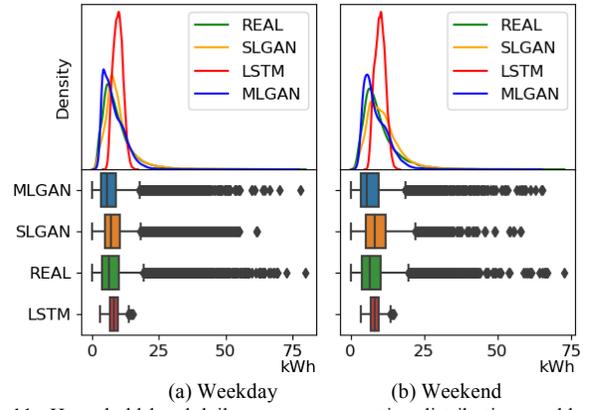

Fig. 11. Household-level daily power consumption distributions and boxplots in weekday and weekend.

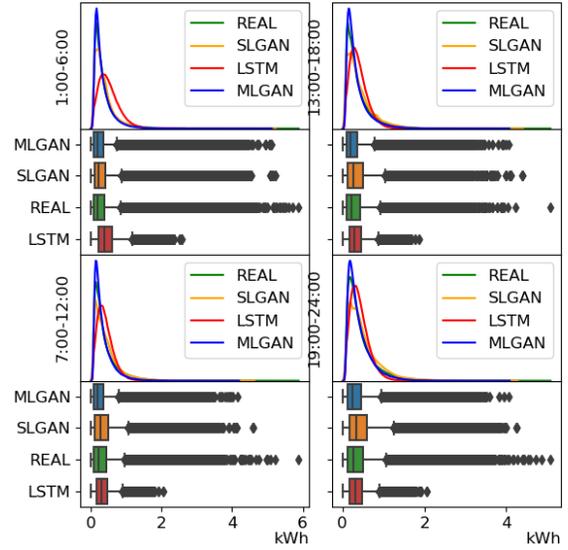

Fig. 12. Household-level hourly power consumption distributions and boxplots in four time periods of a day.

### 2) Evaluation at the Transformer-level

Next, we compare the load group characteristic for the 1424 aggregated profiles in each of the four databases.

- *Mean and peak*. As shown in Fig. 13, all the models have similar performance on the mean value distribution. But SingleLoad-GAN and LSTM tend to generate load groups with higher peak values.

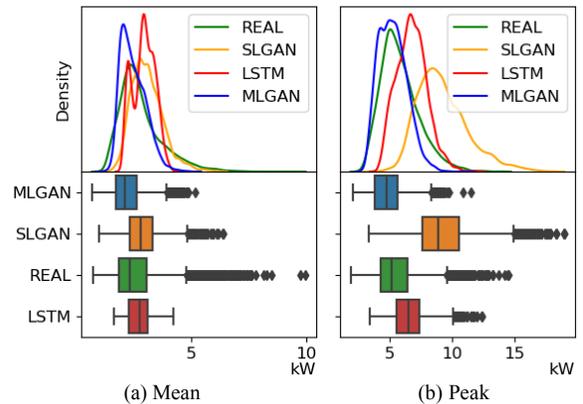

Fig. 13. (a) Transformer-level mean power distributions and boxplots, and (b) Transformer-level peak load value distributions and boxplots.

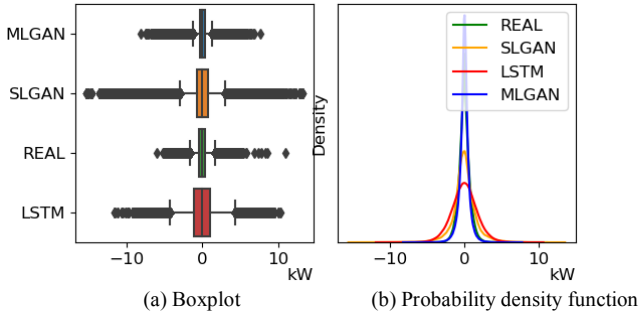

(a) Boxplot        (b) Probability density function

Fig. 14. Transformer-level load ramp distributions and boxplots.

- *Load ramps*. As shown in Fig. 14, MultiLoad-GAN results are smoother than the actual data (i.e., the distribution is more centered towards 0), while SingleLoad-GAN results have more fluctuation. However, LSTM generates more large ramping. Overall, MultiLoad-GAN distribution is closer to the ground truth.
- *Daily and hourly power consumption*. As shown in Figs. 15 and 16, MultiLoad-GAN has comparable performance with the other two methods on daily power consumption and is better than them on hourly power consumption.

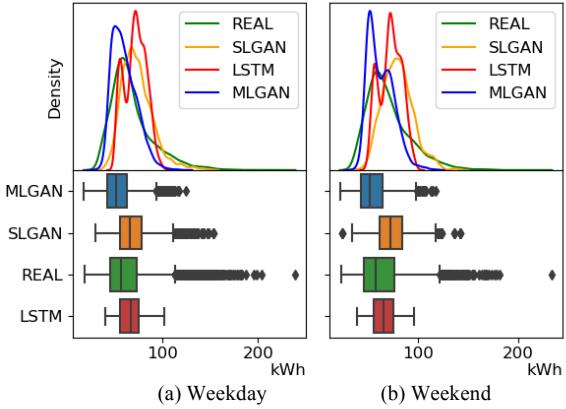

(a) Weekday        (b) Weekend

Fig. 15. Transformer-level daily power consumption distributions and boxplots in weekday and weekend.

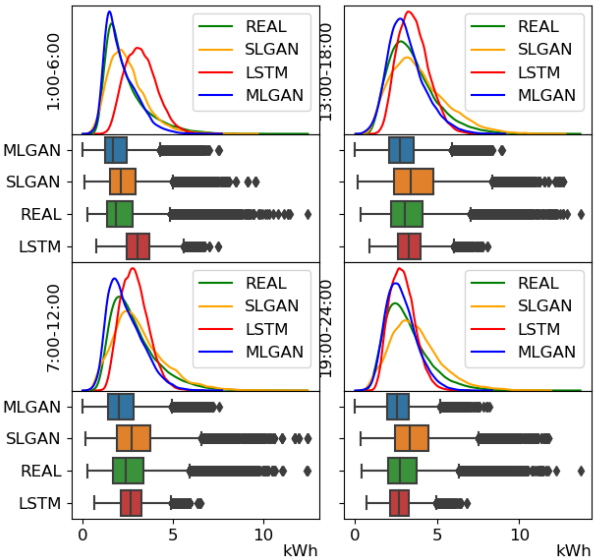

Fig. 16. Transformer-level hourly power consumption distributions and boxplots in four time periods of a day.

To make quantitively comparison, we calculate the FID between the distributions of the generated dataset and real dataset and summarize the results in Table V. We can see that MultiLoad-GAN has comparable performance with SingleLoad-GAN and LSTM on the household-level statistics (3 indices out of 5 perform the best), but show significant advantages on the aggregation-level ones (outperform SingleLoad-GAN and LSTM in all the indices). This means the MultiLoad-GAN can successfully capture correlations between users served by the same transformer. Thus, it can generate load groups with more realistic aggregation-level features while preserving the characteristics of each individual load.

TABLE V
STATISTICAL EVALUATION RESULTS

| Evaluation Criteria | | MultiLoad-GAN | SingleLoad-GAN | LSTM |
|---|---|---|---|---|
| Household Level | Mean | **5.319e-4** | 6.473e-4 | 3.854e-3 |
| | Peak | **2.431e-2** | 4.884e-2 | 9.6.6e-2 |
| | Ramp | 6.700e-4 | **3.445e-4** | 1.650e-1 |
| | Hourly Consumption | (4.125e-4 1.115e-3 4.556e-4 6.726e-4) **6.639e-4** | (1.464e-4 5.098e-4 5.225e-4 1.482e-3) 6.652e-4 | (1.890e-3 2.644e-3 2.200e-3 3.362e-3) 2.524e-3 |
| | Daily Consumption | (3.245e-1 3.609e-1) 3.428e-1 | (2.516e-1 3.411e-1) **2.964e-1** | (2.457 2.561) 4.509 |
| Aggregation Level | Mean | **3.454e-2** | 3.491e-2 | 4.168e-2 |
| | Peak | **5.822e-2** | 2.252 | 1.748e-1 |
| | Ramp | **6.893e-4** | 1.074e-1 | 1.331e-1 |
| | Hourly Consumption | (1.611e-2 3.608e-2 1.728e-2 2.738e-2) **2.421e-2** | (5.634e-3 2.821e-2 3.507e-2 7.370e-2) 3.565e-2 | (9.740e-2 4.661e-2 3.045e-2 6.072e-2) 5.879e-2 |
| | Daily Consumption | (2.030e1 1.943e1) **1.987e1** | (1.649e1 3.230e1) 2.440e1 | (2.490e1 2.195e1) 2.342e1 |

### B. Realisticness Evaluation based on DLC Classification

To train DLC, 4272 generated negative samples and the 1424 "original positive samples" are used as the training set. The positive-negative sample ratio is 1:3. The data set are split into training (80%) and testing (20%) sets.

Three negative sample generation methods are compared: 1) randomly select 8 weekly load profiles from the regional smart meter database; 2) select negative samples based on mean value distribution (see Fig. 5(a)); 3) select negative samples using mean and peak distributions (see Fig. 5, the proposed method).

As shown in Table VI, randomly selecting load profiles as negative samples results in poor identification accuracy, while using the proposed method for NSG, the accuracy can improve to approximately 94%, which is a 20% improvement.

TABLE VI
CLASSIFIER ACCURACY WITH DIFFERENT NSG CRITERIA

| Method No. | Negative sample generation criteria | | Test Accuracy (%) | Mean Confidence Level |
|---|---|---|---|---|
| | Mean | Peak | | |
| DLC | | | 70.37 | |
| | √ | | 92.42 | |
| | √ | √ | **94.38** | **0.9371** |
| RF | √ | √ | 93.77 | 0.8663 |
| K-NN | √ | √ | 90.61 | 0.9254 |
| DT | √ | √ | 89.56 | 1.0 |

To demonstrate the importance of using DLC, several non-DL classifiers, such as Random Forest, K-Nearest Neighbors, Decision Tree, are used for comparison. Negative sample generated based on mean and peak values are used as they have been shown to enhance classifier performance in previous analyses. The result shows that DLC exhibits the highest accuracy and a relative high confidence level. Therefore, we choose DLC as the evaluation model.

The trained DLC will be used to evaluate the realisticness of the load groups generated by MultiLoad-GAN and SingleLoad-GAN. For all four data sets (i.e., real-world load group samples $\mathbf{P}_{672\times8}$, MultiLoad-GAN generated samples $\ddot{\mathbf{P}}_{672\times8}$, SingleLoad-GAN generated samples $\hat{\mathbf{P}}_{672\times8}$, and LSTM forecasted samples). The DLC will give a score (i.e., the confidence level) for each sample to show realisticness. The probability density distribution of the scores are shown in Fig. 17(a). Key statistics are summarized in the first column of Table VII.

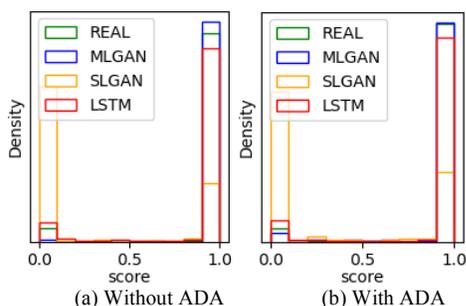

(a) Without ADA      (b) With ADA

Fig. 17. Distribution of DLC scores (a) without ADA and (b) with ADA.

TABLE VII
RESULTS OF DLC-BASED EVALUATION

| Indices | Real data | Single Load-GAN | LSTM | MultiLoad-GAN | MultiLoad-GAN (with ADA) |
|---|---|---|---|---|---|
| Percent of Real | 94.38% | 19.69% | 84.83% | 99.06% | **94.99%** |
| Mean Confidence Level | 0.9371 | 0.1913 | 0.8919 | 0.9899 | **0.9491** |
| Fréchet inception distance with the *Real* load group | N/A | 0.5173 | 0.00706 | 0.01106 | **0.000055** |

From the results, we have the following observations:

- As shown in Fig. 17, DLC is confidence about the classification results, because most scores are close to 1 (real) or 0 (fake).
- As shown in Table VII, 99.06% of the MultiLoad-GAN generated samples are classified as real by DLC, 84.83% of the LSTM forecasted samples are classified as real, while only 19.69% of the SingleLoad-GAN generated samples are classified as real. This means that MultiLoad-GAN generates load groups with similar high-level features with those of the actual load groups.
- The FID index defined in (12) is calculated to measure the similarities between two distributions in Fig. 17(a). The FID between "MultiLoad-GAN" and "Real" is 0.01106, FID between "LSTM" and "Real" is 0.007.58, while between SingleLoad-GAN and "Real" is 0.5173.

This result shows that the MultiLoad-GAN generated load groups are much closer to the ground truth ones from the classifier's viewpoint.

### E. Automatic Data Augmentation

ADA training starts from the MultiLoad-GAN and DLC trained in the previous sections. The loss curves of MultiLoad-GAN in ADA process are shown in Fig. 7(a) stage 2. The performance indices for the with/without ADA-boosted MultiLoad-GAN cases are summarized Table VII (indices are defined in section II. E. 2.) and Fig. 17(b). The results show that the ADA process significantly shorten the distance between MultiLoad-GAN generated data set and the real data set. This shows that the ADA process avoids MultiLoad-GAN to be over-trained so that it only generates load groups strongly resemble the "original positive samples".

Without incorporating the ADA process, the MultiLoad-GAN model generates high-confidence load groups according to the DLC's perspective, resulting in a near 99% real percentage. However, this does not align well with the actual data, where low-confidence load groups exist, leading to a *POR* of 94.38%. By implementing the ADA process, the MultiLoad-GAN model generates additional low-confidence samples, improving its performance to achieve a *POR* of 94.99%, which is closer to the real dataset. Figure 17(b) visually demonstrates the resulting dataset's distribution, which is closer to that of the original dataset. By considering the *POR* before and after the ADA process, it can be concluded that ADA has enhanced the performance of MultiLoad-GAN by approximately 4.07% (moving closer to the real dataset).

## IV. CONCLUSION

In this paper, we present MultiLoad-GAN framework for generating a group of load profiles simultaneously while preserving the spatial-temporal correlations between load profiles in the group. Inspired by the successful application of the GAN-based model in both image processing and power system domain, we develop a novel profile-to-image coding method to convert time-series plots to image patches, so that GAN-based models can be readily used for processing groups of load profiles. To solve the data scarcity problem, we developed an iterative data augmentation process to train MultiLoad-GAN and a classifier alternatively. Thus, the classifier can be used to automatically label positive and negative samples for augmenting the training of both the classifier and the MultiLoad-GAN in subsequent steps. Our simulation results, based on statistical and DLC evaluation, show that compared with the state-of-the-art synthetic load generation process, MultiLoad-GAN better preserves both household-level and group-level load characteristics.

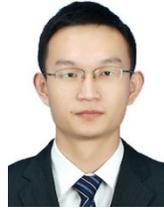

**Yi Hu** (Student Member, IEEE) received the B.S. degree in communication engineering from Chongqing University of Post and Telecommunications, Chongqing, China, in 2014, and the M.S. degree in electrical and communication Engineering from Peking University, Beijing, China, in 2018. Currently he is a Ph.D. candidate in Electrical and Computer Engineering with the Future Renewable Electric Energy Delivery and Management (FREEDM) Systems Center, North Carolina State University, Raleigh, USA. His research interests include synthetic data generation, machine learning and data analytics in power distribution systems.

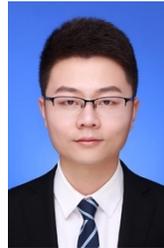

**Yiyan Li** (Member, IEEE) received the B.S. and Ph.D. degrees in electrical engineering from Shanghai Jiao Tong University, Shanghai, China, in 2014 and 2019, respectively. From 2019 to 2022, he was a Postdoctoral Researcher with FREEDM Systems Center, North Carolina State University, Raleigh, USA. He is currently an Assistant Professor with the College of Smart Energy, Shanghai Jiao Tong University. His research interests include machine learning and data analytics in power distribution systems, including synthetic data generation and situational awareness.

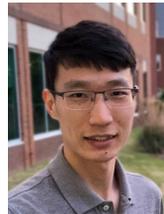

**Lidong Song (**Student Member, IEEE**)** received the B.S. degree in electrical engineering from the China University of Mining and Technology, Beijing, in 2016, and the M.S. degree in electrical engineering from Xi'an Jiao Tong University, in 2019. He is currently pursuing the Ph.D. degree in electrical engineering with North Carolina State University, Raleigh, USA. He is currently working on demand side data analysis, including load profile super-resolution, non-intrusive load monitoring, and baseline load estimation. His research interests include deep learning application and data-driven model development in distribution systems.

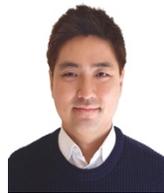

**Han Pyo Lee** (Student Member, IEEE) received the B.S. degree in electrical and electronic engineering from the Yonsei University, Seoul, S.Korea, in 2008, and the M.S. degree in electrical engineering and computer science from the University of Michigan, Ann Arbor, MI, USA, in 2020. He is currently a Ph.D. candidate in electrical and computer engineering with the Future Renewable Electric Energy Delivery and Management (FREEDM) Systems Center, North Carolina State University, Raleigh, NC, USA. His research interests include distribution systems modeling, and combining advances in machine learning and data analytics with the reliability and interpretability inherent in physics-based approaches.

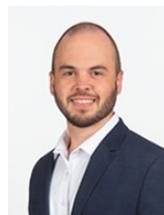

**PJ Rehm** is Vice President, Grid Innovation and Safety. Since joining ElectriCities in July 2012, PJ has held various positions focused on helping member utilities advance and deploy utility technologies. He has helped members with AMI system deployments, demand response programs, renewable programs, customer engagement solutions, and more. PJ also leads ElectriCities' Safety & Training team that provides safety services, education, and support to member utilities. PJ has a Bachelor of Science in mechanical engineering from North Carolina State University.

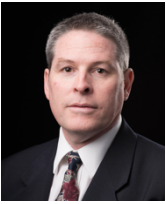

**Matthew Makdad** P.E. has been the Engineering Supervisor at industry partner New River Light & Power in Boone since 2017. He oversees planning and operations which include the line workers, metering shop, and mapping functions. He has a B.S. in Electrical Engineering from The University of Pennsylvania and earned his NC Professional Engineering license in 2003. He has been integral in reviewing the project analysis and performing data field validation.

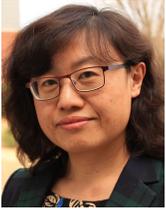

**Ning Lu** (Fellow, IEEE) received the B.S. degree in electrical engineering from the Harbin Institute of Technology, Harbin, China, in 1993, and the M.S. and Ph.D. degrees in electric power engineering from Rensselaer Polytechnic Institute, Troy, NY, USA, in 1999 and 2002, respectively. She is a Professor with Electrical and Computer Engineering Department, North Carolina State University. Her research interests include modeling and control of distributed energy resources, microgrid energy management, and applying machine learning methods in energy forecasting, modeling, and control.